\documentclass[revtex4]{emulateapj}
\usepackage{color}

\def\ltsima{$\; \buildrel < \over \sim \;$}
\def\simlt{\lower.5ex\hbox{\ltsima}}
\def\gtsima{$\; \buildrel > \over \sim \;$}
\def\simgt{\lower.5ex\hbox{\gtsima}}

\shorttitle{Sub-Chandrasekhar mass SNIa}
\shortauthors{McWilliam, Piro, Badenes, \& Bravo}

\begin{document}

\topmargin 0.5in

\newcommand{\znh}{[{\rm Zn/H}]}
\newcommand{\msol}{M_\odot}
\newcommand{\etal}{et al.\ }
\newcommand{\delv}{\Delta v}
\newcommand{\kms}{km~s$^{-1}$ }
\newcommand{\cm}[1]{\, {\rm cm^{#1}}}
\newcommand{\N}[1]{{N({\rm #1})}}
\newcommand{\e}[1]{{\epsilon({\rm #1})}}
\newcommand{\f}[1]{{f_{\rm #1}}}
\newcommand{\rAA}{{\AA \enskip}}
\newcommand{\sci}[1]{{\rm \; \times \; 10^{#1}}}
\newcommand{\ltk}{\left [ \,}
\newcommand{\ltp}{\left ( \,}
\newcommand{\ltb}{\left \{ \,}
\newcommand{\rtk}{\, \right  ] }
\newcommand{\rtp}{\, \right  ) }
\newcommand{\rtb}{\, \right \} }
\newcommand{\ohf}{{1 \over 2}}
\newcommand{\nohf}{{-1 \over 2}}
\newcommand{\rhf}{{3 \over 2}}
\newcommand{\smm}{\sum\limits}
\newcommand{\perd}{\;\;\; .}
\newcommand{\cmma}{\;\;\; ,}
\newcommand{\intl}{\int\limits}
\newcommand{\mkms}{{\rm \; km\;s^{-1}}}
\newcommand{\ew}{W_\lambda}


%
%
\title{Evidence for a Sub-Chandrasekhar Mass Type~Ia 
Supernova in the Ursa Minor Dwarf Galaxy}

\author{Andrew McWilliam$^1$ and Anthony L. Piro$^2$}
\affil{The Observatories of the Carnegie Institution for Science, \\
813 Santa Barbara Street, Pasadena, CA 91101}
\email{$^1$ andy@carnegiescience.edu} 
\email{$^2$ piro@carnegiescience.edu} 

\author{Carles Badenes$^3$}
\affil{Department of Physics and Astronomy, University of Pittsburgh,\\
3941 O'Hara Street, Pittsburgh, PA 15260}
\email{$^3$ badenes@pitt.edu}

\author{Eduardo Bravo$^4$}
\affil{E.T.S. Arquitectura del Vall\'es, Universitat Polit\'ecnica de Catalunya, 
Carrer Pere Serra 1-15, E-08173 Sant Cugat del Vall\'es, Spain}
\email{$^4$ eduardo.bravo@upc.edu}

\begin{abstract}
A longstanding problem is identifying the elusive progenitors of Type Ia 
supernovae (SNe Ia), which can roughly be split into Chandraksekhar and 
sub-Chandrasekhar mass events. An important difference between these two cases 
is the nucleosynthetic yield, which is altered by the increased neutron excess in 
Chandrasekhar progenitors due to their pre-explosion simmering and high 
central density. Based on these arguments, we show that the chemical 
composition of the most metal-rich star in the Ursa Minor dwarf galaxy, 
COS\,171, is dominated by nucleosynthesis from a low-metallicity, low-mass, 
sub-Chandrasekhar mass SN Ia. Key diagnostic abundance ratios include Mn/Fe 
and Ni/Fe, which could not have been produced by a Chandrasekhar-mass SN Ia. 
Strong deficiencies of Ni/Fe, Cu/Fe and Zn/Fe also suggest the absence of 
alpha-rich freeze-out nucleosynthesis, favoring low-mass WD progenitor SN Ia, 
near $0.95\,M_\odot$ from comparisons to numerical detonation models. We also 
compare Mn/Fe and Ni/Fe ratios to the recent yields predicted by Shen et al., 
finding consistent results. To explain the [Fe/H] at $-$1.35 dex for COS\,171 
would require dilution of the SN Ia ejecta with $\sim10^4\,M_\odot$ of 
material, which is expected for a SN remnant expanding into a warm 
interstellar medium with $n\sim1\,{\rm cm^{−3}}$. In the future, finding more 
stars with the unique chemical signatures we highlight here will be important 
for constraining the rate and environments of sub-Chandrasekhar SNe Ia.

\end{abstract}

%
%
%
%
%



\section{Introduction}

The low average metallicity of most dwarf galaxies suggests that, like the
Milky Way (MW) halo, star formation (SF) was truncated in these 
systems prior to complete gas consumption, presumably due to
gas loss (e.g., Hartwick 1976).  Thus, dwarf galaxies offer the 
potential to study the early phases of chemical enrichment.

In the Type~Ia supernova (henceforth SNIa) time-delay picture of chemical
evolution (e.g., Matteucci \& Brocato 1990; henceforth MB90) dwarf galaxies 
experience a low specific star formation rate (SFR), resulting in an
increased fraction of nucleosynthesis products from SNIa versus
core-collapse, Type~II, supernovae (henceforth SNII), compared to the 
MW at any given metallicity.  In this scenario MB90 predicted that dwarf
galaxies would show low [$\alpha$/Fe] ratios, compared to the MW disk, due 
to enhanced iron production from SNIa without extra $\alpha$-element 
synthesis from SNII.  These low [$\alpha$/Fe] ratios were subsequently 
observed (e.g., Shetrone et al. 2001, 2003).

Thus, it appears that dwarf galaxies are enhanced in SNIa ejecta;
so, the chemical composition of stars in these systems may be employed
as probes of SNIa nucleosynthesis and thereby provide constraints on the
SNIa mechanism.

For recent reviews of SNIa scenarios and variants, including
nucleosynthesis predictions, see Seitenzahl \& Townsley (2017) and
Maoz et al. (2014).  One
long-standing scenario for SNIa involves an explosion following transfer
of mass from a companion onto a white dwarf (WD) near the Chandrasekhar
mass limit (e.g., Whelan \& Iben 1973).  

As first noted by Arnett (1971)
iron-peak nucleosynthesis depends strongly on the neutron excess
\footnote{$\eta$=(N-Z)/(N+Z) where N and Z are neutron and proton 
numbers respectively}, $\eta$, during the explosion;
in particular, the yields for neutron-rich species like $^{51}$V, $^{55}$Mn
and $^{58}$Ni are reduced at low neutron excess values.  

Timmes et al. (2003) suggested that a dispersion of SNIa metallicities
could be responsible for much of the intrinsic variation in the 
luminosity of SNIa, due to the dependence of explosively-produced
$^{56}$Ni on the neutron excess, which changes markedly with metallicity.

Detailed nucleosynthesis calculations for Chandrasekhar-mass SNIa scenarios 
(e.g., Piro \& Bildsten 2008; Chamulak et al. 2008;
Mart\'inez-Rodr\'iguez et al. 2016; and Piersanti et al. 2017)
showed that for $\sim$1,000 years prior to the SNIa explosion low-level
carbon burning, or {\em simmering}, occurs which increases $\eta$ to a
value roughly equivalent to half solar composition.  
In particular, the Piersanti et al. (2017) calculations show that very low 
metallicity Chandrasekhar-mass SNIa have a floor in $\eta$ at the time of
explosion near $\eta$=6.7$\times$10$^{-4}$; however, at higher metallicity, 
SNIa show correspondingly larger $\eta$. Thus, even with the
$\eta$ increase due to simmering, SNIa nucleosynthesis yields reflect
their original metallicity.   However, the basic conclusion from 
these studies is that all Chandrasekhar-mass SNIa experience pre-explosion 
simmering with an increase to high $\eta$.

Thanks to this increased $\eta$, even quite metal-poor Chandrasekhar-mass
WD produce SNIa Mn/Fe and Ni/Fe yield ratios that are not too different
from the solar values.

Alternate SNIa scenarios (e.g., Iben \& Tutukov 1984; Webbink 1984; 
Fink et al. 2007)
involves detonation following rapid helium accretion, or a violent
collision, or merger, of two sub-Chandrasekhar mass WDs, ultimately
triggering a detonation in the primary. Importantly, the simmering phase 
does not occur in these sub-Chandrasekhar mass models, so there can be 
no increase in $\eta$ beyond that provided by the primordial metallicity,
unlike the Chandrasekhar-mass models.  Furthermore, according to
Seitenzhal et al. (2013) the critical density required to produce $^{55}$Mn
following normal freeze-out after nuclear statistical equilibrium (NSE)
($\rho$$\geq$$2\times 10^8$ g cm$^{-3}$) is not reached for WD masses 
below $\sim$1.2M$_{\odot}$, suggesting no Mn production in sub-Chandrasekhar
mass models.  Contrary to this assertion, however, the 0.88 to 1.15M$_{\odot}$ 
sub-Chandrasekhar mass
SNIa models of E.~Bravo (introduced in Yamaguchi et al.  2015) show a factor
of 10 range in both Mn/Fe and Ni/Fe yield ratios, depending on
initial mass and metallicity; these results are confirmed in the
recent work of Shen et al. (2017).

Notwithstanding these details, low-metallicity sub-Chandrasekhar mass
SNIa are expected to produce very low Mn/Fe and Ni/Fe
ratios, distinctly lower than the near-solar values expected
from Chandrasekhar-mass SNIa.  Thus, the Mn/Fe and Ni/Fe yield ratios
are intimately related to the SNIa mechanism and may be employed for
diagnostic purposes.


While direct measurement of iron-peak element ratios ratios in 
supernova remnants (e.g. Badenes et al. 2008a; Yamaguchi et al. 2015;
Mart\'inez-Rodr\'iguez et al. 2017) 
provides a way to probe the SNIa mechanism (e.g., Bravo 2013),
the important role of 
SNIa in the chemical evolution of dwarf galaxies suggests that the
composition of these systems may also be of use (e.g., North et al. 2013;
Kobayashi, Nomoto \& Hachisu 2015).
We may expect
chemical signatures from SNIa to be enhanced in low-metallicity, 
low-mass, dwarf galaxies, where the chemical enrichment by SNII is
truncated and small numbers of SNe could potentially have a significant
and measurable effect on chemical composition.  Based on
progenitor lifetimes, the order of iron-peak nucleosynthesis might
reasonably be: SNII, followed by Chandrasekhar mass SNIa, and
finally sub-Chandrasekhar mass SNIa.  Thus, one might expect that the 
most metal-rich stars
in a dwarf galaxy are more likely to result from a sub-Chandrasekhar mass
SNIa phase of chemical enrichment.


This work was motivated by the unusual composition of the most metal-rich 
star, COS\,171, in the Ursa Minor dwarf galaxy (henceforth UMi), as measured
by Cohen \& Huang (2010; henceforth CH10).
We compare the chemical composition of COS\,171 to other UMi stars and
the Milky Way halo.  Accordingly, we identify a chemically normal star,
UMi~28104, useful as a standard to isolate the composition of the
contamination event that
produced COS\,171.  We confirm the published LTE abundance calculations of CH10
and, when possible, we have applied differential non-LTE corrections.  After
comparison of our final chemical abundance ratios with predicted yields from
a variety of supernova nucleosynthesis scenarios, we conclude that the
COS\,171 composition resulted from a low-mass, metal-poor, sub-Chandrasekhar 
mass SNIa event.  In a chemical evolution context, this is most easily
understood as due to a sub-Chandrasekhar mass SNIa diluted with
$\sim$10$^4$ M$_{\odot}$ of hydrogen, consistent with expectations for
a supernova remnant expanding into a warm interstellar medium.


\section{The Chemical Composition of Ursa Minor}

The detailed chemical composition of stars in UMi
has been investigated by Shetrone, C\^ote \& Sargent (2001),
Sadakane et al. (2004), and CH10.

The seminal work of Shetrone et al. (2001) found
that the chemical compositions of UMi, Draco and Sextans dSphs
are characterized by a large dispersion in [Fe/H], low [$\alpha$/Fe]
ratios, low [Y/Fe], and [Ba/Eu] ratios indicating r-process 
nucleosynthesis.  Thus, the UMi chemical composition is distinct from
the MW halo.

Sadakane et al. (2004) confirmed the low [$\alpha$/Fe] ratios in UMi,
and, for one star near [Fe/H]=$-$1.5, found very strong over-abundances
of neutron-capture elements, matching the solar system r-process pattern.

The most extensive detailed chemical composition study of UMi was undertaken 
by CH10, who examined 10 UMi RGB stars.  They found a range in [Fe/H] from 
$-$3.10 to $-$1.35 dex, roughly normal O/Fe, Mg/Fe, and Si/Fe
for the most metal-poor UMi stars, but generally declining [$\alpha$/Fe]
with increasing metallicity, well below the MW halo trend, particularly for
stars above [Fe/H]$\sim$$-$2.  Critically, three of the four most metal-rich
UMi stars show [Eu/Fe] above $+$0.60 dex (but up to $+$0.87 dex), and
all 4 stars above [Fe/H]=$-$1.9 show heavy element ratios consistent 
with pure r-process composition.  In this way, UMi is similar to the
r-process dwarf galaxy Reticulum~II (Ji et al. 2016), although Ret~II 
has higher [$r$/Fe].

\subsection{The Unusual UMi Star COS\,171}

At [Fe/H]=$-$1.35 dex, the most metal-rich star in the CH10 sample,
COS~171, has an extraordinary chemical composition, including: strong
deficiencies (exceeding 0.6 dex) of [C/Fe], [Na/Fe], [Sc/Fe], [V/Fe],
[Mn/Fe], [Ni/Fe], [Cu/Fe] and [Zn/Fe] ratios, as well as sub-solar
ratios, below $-$0.3 dex, for [Mg/Fe], [Ca/Fe], [Ti/Fe], [Cr/Fe], 
and [Co/Fe].

In the following discussion of COS\,171, we consider only the UMi
abundance results of CH10, in order to avoid complications from 
systematic measurement differences between studies.

\begin{figure}[h]
\centering
\includegraphics[width=8.0cm]{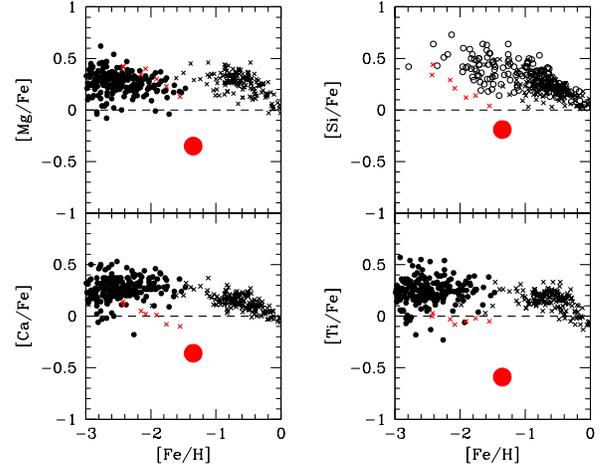}
\caption{[$\alpha$/Fe] ratios in UMi (red) compared to the MW (black) halo,
thin, and thick
disks.  The large filled red circle indicates UMi star COS171, while small
red crosses show other UMi stars from Cohen \& Huang (2010).  Black crosses 
indicate MW thin and thick disk stars, and some halo stars, from Reddy et al. 
(2006); black filled circles are MW metal-poor halo stars from Barklem et al. 
(2005); black open circles show [Si/Fe] for MW halo and disk stars from Fulbright 
(2000).  In the standard chemical evolution paradigm of Matteucci \& Brocato
(1990), the low [$\alpha$/Fe] ratios suggest heavy contamination by SNIa
ejecta in UMi, compared to the MW, increasing with increasing [Fe/H].
}
\label{fig-ch10-alphafe}
\end{figure}

Figure~\ref{fig-ch10-alphafe} shows [X/Fe] versus [Fe/H] in UMi for the 
sample of stars studied by CH10, with small red crosses, compared to
the MW halo, thick and thin disks (black symbols).  It is immediately obvious
that UMi shows depleted [$\alpha$/Fe] ratios, with a general trend that
is qualitatively consistent with the scenario of MB90,
in which dwarf galaxies are predicted to show the decline in [$\alpha$/Fe]
at lower [Fe/H] than the MW due to a reduced SFR.  Since
this reduction of [$\alpha$/Fe] is thought to be due to the contribution of
iron from SNIa, the UMi stars appear to show an increasing, and relatively
large, SNIa/SNII ratio.  Notably, the most iron-rich star in UMi,
COS\,171 (shown as the large filled red circle), exhibits extraordinarily low
[$\alpha$/Fe] ratios compared to any MW study.

At closer inspection, Figure~\ref{fig-ch10-alphafe} shows that, excluding the
most metal-poor star in UMi, the [Mg/Fe] ratios in UMi are not as depleted as
the [Si/Fe], [Ca/Fe] and [Ti/Fe] ratios; indeed, the [Mg/Fe] ratios appear
normal compared to the MW halo (again, except for the most metal-rich UMi 
star).  If [$\alpha$/Fe] deficiencies in dwarf galaxies are supposed to
be due to a reduced SFR, as predicted by MB90, resulting from excess
SNIa iron at low [Fe/H], then the halo-like [Mg/Fe] in UMi stars is
unexpected: naively, all [$\alpha$/Fe] ratios should decline together.

The implied [Mg/Ca] enhancement, near $+$0.3--0.4 dex, might be explained by
an over-representation of massive SNII.  Much larger [Mg/Ca] ratios,
near $+$0.95 dex, have been seen in the Hercules dwarf (henceforth Her) by
Koch et al. (2008). Because Mg production is made almost exclusively by
massive core-collapse, SNII, events with progenitor masses exceeding
$\sim$30 M$_{\odot}$ (e.g. Woosley \& Weaver 1995), the high [Mg/Ca] ratios
in Her indicate enhanced pollution by high-mass SNII events.  Koch
et al. (2008) suggested that for Her this could result from stochastic 
sampling of the initial mass function (henceforth IMF) that,
by chance, favored massive stars.  However, the probability of randomly 
selecting only massive stars, above 30M$_{\odot}$, from the IMF diminishes 
rapidly with the number of samplings.  Such a mechanism could only
occur if a small number of SNII events (fewer than 11) produced the Her
chemical composition.  Similarly, the enhanced [Mg/Ca] ratios in UMi may
be the signature of stochastic sampling of the SNII IMF in the chemical
evolution of this dwarf galaxy.

We note that in Figure~\ref{fig-ch10-alphafe} the [Mg/Fe] ratio for COS\,171
is markedly lower than would be extrapolated from the trend at lower [Fe/H]. 
Indeed, while the bulk of UMi stars show [Mg/Ca]$\sim$$+$0.3 dex, for 
COS\,171 [Mg/Ca]=$+$0.0 dex.  This suggests some production of Ca
in the composition of COS\,171.

\begin{figure}[h]
\centering
\includegraphics[width=8.0cm]{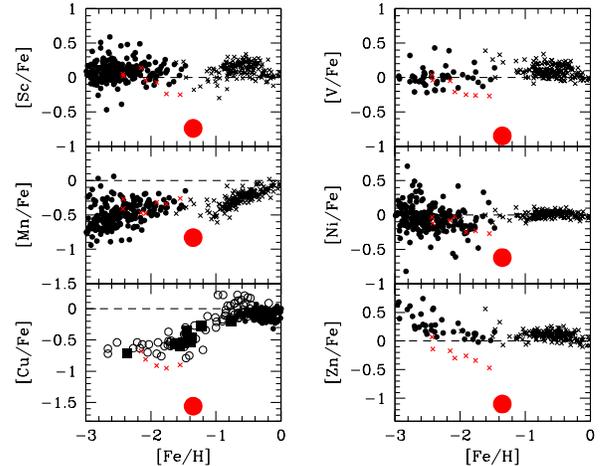}
\caption{Iron-peak [X/Fe] ratios in UMi compared to the MW halo, thin, and
thick disks.  Symbols are the same as in Fig.~\ref{fig-ch10-alphafe}.  
COS171 has among the lowest [X/Fe] ratios reported for the odd-numbered
elements Sc, V, Mn and Cu.
}
\label{fig-ch10-iron-peak}
\end{figure}

The iron-peak elements in Figure~\ref{fig-ch10-iron-peak} show generally good
agreement with the MW stars, but severe under abundances for the most metal-rich
UMi star, COS171.  In particular, the odd-numbered elements, Sc, V, Mn, and
Cu are neutron rich, so their deficiency suggests
a rather low neutron excess, $\eta$, (or equivalently high electron fraction,
$Y_e$).

\begin{figure}[h]
\centering
\includegraphics[width=8.0cm]{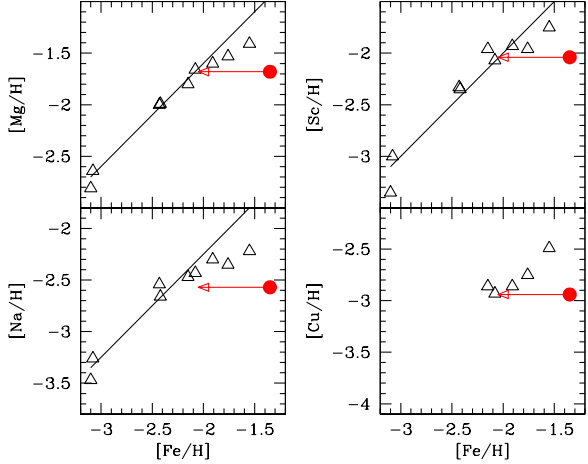}
\caption{ [X/H] versus [Fe/H] for Mg, Na, Sc and Cu in UMi, reported by
Cohen \& Huang (2010).  For these elements, the composition of the highest
[Fe/H] star, COS171 (filled red circle), more closely resembles UMi stars 
near [Fe/H]=$-$2.0 dex.
The red arrow indicates the effect of a reduction in [Fe/H] by 0.7 dex, which
suggests that the COS171 composition resulted from the addition of
0.7 dex of iron-peak material to a pre-existing mixture.
The black line for Mg, Na and Sc indicates a 1:1 between [X/H] and [Fe/H].
}
\label{fig-xfe1-vs-feh}
\end{figure}

\begin{figure}[h]
\centering
\includegraphics[width=8.0cm]{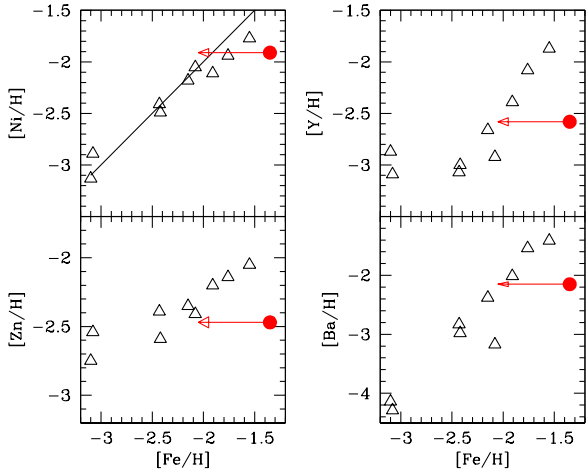}
\caption{ [X/H] versus [Fe/H] for Ni, Zn, Y and Ba in UMi, reported by
Cohen \& Huang (2010).  The composition of the highest
[Fe/H] star, COS171 (filled red circle), more closely resembles UMi
stars near [Fe/H]=$-$2.0 dex,
even for the neutron-capture elements, but with an [Fe/H] enhancement
near 0.7 dex.  The red arrow indicates the effect of a reduction in [Fe/H] 
by 0.7 dex.  The black line indicates a 1:1 relation between [Ni/H] and 
[Fe/H].
}
\label{fig-xfe2-vs-feh}
\end{figure}

It occurs to us that the excessively low [X/Fe] ratios for the 8 elements 
in COS\,171, shown in Figures~\ref{fig-ch10-alphafe} and 
\ref{fig-ch10-iron-peak}, might 
be largely explained by the addition of nearly pure iron to a pre-existing 
composition.  In Figures~\ref{fig-xfe1-vs-feh} and \ref{fig-xfe2-vs-feh} we 
show the UMi [X/H] versus [Fe/H] LTE abundances from CH10; some panels
show a 1:1 line.   It is clear from these figures that a simple shift to
lower [Fe/H] by $\sim$0.7 dex would bring the COS\,171 [X/H] 
ratios into approximate consistency with the more metal-poor members of
the galaxy, at least for Na, Mg, Sc, Cu, Ni, Zn, Y and Ba.

Figures~\ref{fig-xfe1-vs-feh} and \ref{fig-xfe2-vs-feh}, suggest that
COS\,171 resulted from $\sim$0.7 dex of Fe added to the pre-existing 
galaxy composition.  Given the COS\,171 [Fe/H]=$-$1.35 dex, 
a pre-existing composition of [Fe/H]$\sim$$-$2.05 dex is suggested.   
The CH10 star with metallicity closest to our putative pre-existing
composition is UMi~28104, at [Fe/H]=$-$2.08 dex. 
Because UMi~28104 has a composition similar to the more metal-poor stars
in UMi, with abundance ratios fairly typical of dwarf galaxy stars in
other systems (e.g., Shetrone et al. 2001), we take the chemical composition 
of UMi~28104 to indicate that of UMi just prior to the enrichment
event that produced COS\,171.  




\begin{figure}[h]
\centering
\includegraphics[width=8.0cm]{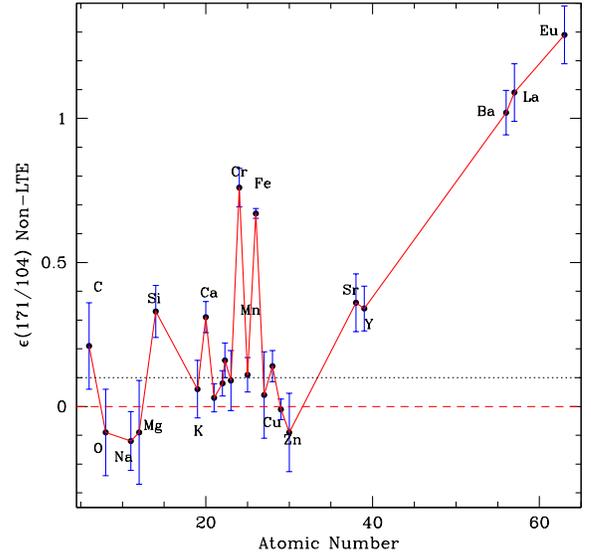}
\caption{The abundance distribution ratio (difference in the log),
$\Delta$$\varepsilon$(171$-$28104), including non-LTE corrections
(see section~\ref{sec-nlte}).  The unusual composition of UMi
star \#171 is evident.  This figure includes revised abundances for
O and K, based on our re-analysis of the CH10 spectra.
}
\label{fig-dabund}
\end{figure}

Figure~\ref{fig-dabund} shows the abundance ratios of COS\,171 relative to
the standard, UMi~28104 (i.e. this is the difference in log abundances);
we emphasize that this is {\bf not} the difference in the number of atoms.

The abundance ratios in Figure~\ref{fig-dabund} not
only highlight the enhanced elements in COS\,171, but thanks to the similar
stellar atmosphere parameters of UMi~28104, taking abundance ratios mitigates
a number of systematic measurement errors (e.g., gf values,
model atmosphere errors, non-LTE effects, temperature scale errors, etc).
While the abundance ratios in Figure~\ref{fig-dabund} are based on the
CH10 LTE abundances, the points in the figure have been corrected for 
differential non-LTE effects, whenever possible (see section~\ref{sec-nlte}).  

Error bars in Figure~\ref{fig-dabund} were taken from
the abundance dispersions given by CH10, or in the case of oxygen,
from the EW uncertainties.
The main difference between the ratios presented in Figure~\ref{fig-dabund}
compared to the CH10 LTE results is for Mn: with non-LTE corrections the Mn
enhancement in COS\,171, at $+$0.11 dex, is less than 1$\sigma$ from zero,
whereas the uncorrected CH10 result is $+$0.38 dex.
In Figure~\ref{fig-dabund} the displayed oxygen abundance ratio is not from
CH10, but indicates the value found in this work
(see section~\ref{sec-reanalysis})
from the original CH10 spectra.  Thus, we find no oxygen enhancement for
COS\,171 relative to UMi~28104, wheres the original CH10 results suggested
a higher oxygen abundance for COS\,171 by 0.31 dex.

Figure~\ref{fig-dabund} shows strong enhancements of neutron-capture
elements (i.e., elements heavier than zinc), increasing with increasing
atomic number, which appears to be due to an r-process enrichment event that
affected all UMi stars above 
[Fe/H]$\sim$$-$1.9 dex.  Since three such r-process rich stars do not share
the unusual iron-peak composition of COS\,171, we assume that the r-process
enrichment event was an unrelated phenomenon.

Excluding the neutron-capture elements, Figure~\ref{fig-dabund} shows clear
detection of Si, Ca, Cr and Fe enhancements in COS\,171, well above the
individual measurement uncertainties.  Marginal detections, comparable to
the estimated error bars, occur for C, K\footnote{Based on our re-analysis of
the K abundance for COS\,171}, Ti, V, Mn, and Ni, while O, Na, Mg,
Sc, Co, Cu and Zn show no evidence for enhancement.  


\subsection{A Check on the LTE Results}

As a check on CH10 results, for UMi~28104 and the unusual UMi star
COS171 we have employed the CH10 equivalent widths (EWs) and atmosphere
parameters to compute element abundances, using the 2014 version of
the LTE spectrum synthesis program MOOG (Sneden 1973).  Our 
abundances are in very good agreement with CH10, including
the excitation temperatures derived from Fe~I lines, thus supporting the
unusual composition for COS171 claimed by CH10.


\subsection{Oxygen and Potassium Re-analysis}
\label{sec-reanalysis}

The CH10 abundance of oxygen in COS~171 was based on only two lines:
[O~I] at 6363\AA\ and O~I at 7771\AA , with putative EWs of 12.6 m\AA\ and 
8.5 m\AA ; for UMi~28104 only the 6363\AA\ line was measured, with an EW
of 8.8m\AA .  These EWs are strongly affected by noise, given the 1$\sigma$
EW uncertainty of $\sim$4m\AA , which explains the large abundance difference
derived from the two lines in COS~171.


Curiously, the [O~I] line at 6300\AA , which has a larger $gf$
value than the 6363\AA\ line (roughly three times stronger) was not 
employed by CH10.

In our comparisons of observed and predicted yields from a variety of
nucleosynthesis sites,
the CH10 oxygen abundance showed significant discordance with abundances of 
other elements, such as Mg.  We therefore elected to re-analyze the original 
CH10 Keck/HIRES spectra of COS~171 and UMI~28104, which we downloaded from 
the Keck Archive\footnote{https://www2.keck.hawaii.edu/koa/public/koa.php}.
Reduction to 1-dimensional 
spectra was performed using the program {\em makee} and 
EWs were measured with the IRAF {\em splot} routine.

These spectra show that the 6363\AA\ [O~I] and O~I 7771\AA\ lines are, indeed,
extremely noisy in COS~171 and UMi~28104.  The [O~I] line at 6300\AA\ 
has significant EW and should be readily measured; however, in the CH10
UMi spectra this line is blended with a comparable telluric O$_2$ 
inter-combination (P7) line from the (2-0) $b$--X $^1\Sigma_g^+$--$^3\Sigma_g^-$ 
electronic vibration-rotation transition. This telluric blend suggests why
CH10 did not employ the [O~I] 6300\AA\ line.  In UMi~28104 the
[O~I]/O$_2$ blend at 6300\AA\  has a total EW of 42.8m\AA , while for COS~171
the feature has a total EW of 35.8m\AA .  We estimated the strength of
the blending telluric O$_2$ line from the EW of its spin-doublet, 
located 0.78\AA\ redward.
The two P7 spin-doublet lines have very similar, but not identical, EWs in
telluric standard spectra of hot stars, and the redder of the doublet
lines is unblended in the UMi stellar spectra.
We computed the strength ratio of the O$_2$ line pair from 
a high S/N spectrum of a telluric standard B star, which then enabled 
the strength of the O$_2$ contamination of the [O~I] 6300\AA\ feature 
to be computed from the unblended O$_2$ line. 

For UMi~28104 the telluric-corrected [O~I] 6300\AA\ EW is 21.8m\AA , while for
COS~171 the line is weaker, at 12.1 m\AA .  Incidentally, these values suggest 
6363\AA\ [O~I] line EWs of 7 and 4 m\AA , respectively (compared to the
CH10 values of 8.8 and 12.6m\AA ). Thus, the 6363\AA\ lines are
too weak compared to the measurement uncertainty, of 1$\sigma$$\sim$4m\AA ,
to allow reliable oxygen abundance measurement.  For this reason, we rely 
only on the [O~I] line at 6300\AA\ for oxygen abundances in the UMi stars.

Oxygen abundances were computed here from our telluric-corrected EWs
for the [O~I] 6300\AA\ line, using
the spectrum synthesis program MOOG (Sneden 1973), the stellar model 
atmosphere grid of Kurucz\footnote{http://kurucz.harvard.edu/grids}
and the atmosphere parameters given in CH10.  We find
[O/H] of $-$1.616 and $-$1.712 dex for UMi~28104 and COS~171 respectively,
on the meteoritic solar abundance scale of Asplund (2009).  Curiously, 
COS~171 has a lower [O/H], by 0.10 dex, despite its significantly higher
[Fe/H], but since this oxygen abundance difference is within the
measurement uncertainty of 0.15 dex, the two stars might reasonably have the 
same oxygen abundance.

We also employed the CH10 spectra to check the measured EW for critical
species, including potassium, for which only the K~I line at 7699.0\AA\
was used to determine potassium abundances.  We measured a K~I line EW of
105m\AA\ for COS\,171, significantly smaller than the 130m\AA\ in CH10;
however, we confirmed the CH10 K~I line EW for UMi~28104, at 109m\AA .
Our K~I EW measurement resulted in a decreased potassium abundance for
COS\,171, at $\varepsilon$(K)=3.28 dex, which is within 0.01 dex of the
value found for UMi~28104.  Thus, there appears to have been no measurable
potassium production by the progenitor event to COS\,171.

\subsection{Non-LTE Effects and Abundance Uncertainties}
\label{sec-nlte}

Our use of star UMi~28104 as a comparison to COS\,171 both
accounts for the pre-existing UMi composition present in COS\,171
and eliminates constant systematic effects, such as log~gf
scales.

\begin{deluxetable}{rllr}
\label{tab-cnlte}
\tablecaption{Adopted non-LTE corrections}
\tabletypesize{\scriptsize}
\tablewidth{8.0cm}
\tablehead{\\
\colhead{Ion} &
\colhead{COS\,171} &
\colhead{UMi~28104} &
\colhead{Notes} \\
}
\startdata
Na~I  &   \llap{$-$}0.00    &    \llap{$-$}0.02   &        1   \\
Mg~I  &   \llap{$-$}0.06    &    \llap{$+$}0.01   &        2   \\
Si~I  &   \llap{$-$}0.02    &    \llap{$-$}0.02   &        3   \\
 K~I  &   \llap{$-$}0.30    &    \llap{$-$}0.37   &        4   \\
Ca~I  &   \llap{$+$}0.10    &    \llap{$+$}0.15   &        5   \\
Sc~II &             0.00    &              0.00   &        6   \\
Ti~I  &   \llap{$+$}0.24    &    \llap{$+$}0.39   &        7   \\
Ti~II &             0.00    &              0.00   &        8   \\
 V~I  &             ...     &              ...    &        9   \\
Cr~I  &   \llap{$+$}0.04:   &    \llap{$+$}0.04:  &       10   \\
Mn~I  &   \llap{$+$}0.15    &    \llap{$+$}0.42   &       11   \\
Fe~I  &   \llap{$+$}0.05    &    \llap{$+$}0.11   &       12   \\
Co~I  &   \llap{$+$}0.19    &    \llap{$+$}0.42   &       13   \\
Ni~I  &             0.00    &              0.00   &       14   \\
Cu~I  &             0.2:    &              0.2:   &       15   \\
Zn~I  &   \llap{$-$}0.13    &    \llap{$-$}0.10   &       16   \\
\enddata
\tablecomments{
When possible, the non-LTE corrections are taken from the MPIA non-LTE web page 
(http://nlte.mpia.de/) of Bergemann, which was used for O, Mg, Si, Ti, Mn, Fe and Co.  
For Na we employed non-LTE web page INSPECT, http://inspect.coolstars19.com/, of Lind.
For Ca we used the non-LTE web page of Mashonkina, http://spectrum.inasan.ru/nLTE/.
Source references are cited in these web pages, but are not always correct, or may not
contain information given by the web page.
}
\end{deluxetable}
\vbox{\scriptsize
\noindent{\bf 1.} {Average Na I corrections are -0.057 and -0.087 for 171
    and 104 if Na D lines included}

\noindent{\bf 2.} {See Bergemann et al. (2016).  Also Osorio \& Barklem (2016)
    and Mashonkina (2013)}

\noindent{\bf 3.} {Si corrections in the MPIA web page are not given in the 
    identified reference.}

\noindent{\bf 4.} {Non-LTE corrections for the K~I line at 7699\AA\ 
    estimated from Ivanova \& Shimanski\u{i} 2000}

\noindent{\bf 5.} {Ca non-LTE corrections given by 
    Mashonkina, Sitnova, \& Pakhomov (2016)}

\noindent{\bf 6.} {Sc II non-LTE corrections not found, but assumed to be 
    zero or very small.}

\noindent{\bf 7.} {For Ti I source reference is Bergemann (2011); but see 
    also Mashonkina et al. (2016).}

\noindent{\bf 8.} {Ti II non-LTE corrections Mashonkina et al. (2016) are 
    negligibly small.}

\noindent{\bf 9.} {We could not find non-LTE corrections for vanadium.}

\noindent{\bf 10.}{Lawler et al. (2017) found non-LTE corrections for Cr 
    near $+$0.04 dex in the metal-poor dwarf HD84937. 
    Bergemann \& Cescutti (2010) also presented Cr non-LTE corrections 
    for dwarf stars.}

\noindent{\bf 11.}{MPIA web page cite Mn non-LTE corrections from 
    Bergemann \& Gehren (2008)}

\noindent{\bf 12.}{MPIA web page cite Fe non-LTE corrections from 
    Bergemann et al. (2012)}

\noindent{\bf 13.}{MPIA web page cited cobalt non-LTE corrections from 
    Bergemann et al. (2010)}

\noindent{\bf 14.}{Wood et al. (2014) found no evidence of non-LTE effect 
    on Ni in the metal-poor dwarf HD84937.}

\noindent{\bf 15.}{Our estimated Cu non-LTE corrections are based on 
    calculations for metal-poor dwarf stars by Yan et al. (2015,2016);
    see also Shi et al. (2014).}

\noindent{\bf 16.}{Zinc non-LTE corrections estimated from metal-poor 
    dwarf calculations by Takeda et al (2005), who assumed no collisions 
    with hydrogen, S$_{\rm H}$=0.}
}

One important correction enabled by taking differences between
COS\,171 and UMi~28014 is that these stars are both metal-poor RGB stars
and both likely suffer similar non-LTE effects on the derived LTE abundances.
Thus, to first-order the non-LTE effects cancel-out when comparing 
COS\,171 to UMI~28104, but because the latter is more metal-poor,
by $\sim$0.7 dex, differential non-LTE effects are possible.
In particular,
non-LTE effects are typically more extreme in more metal-poor 
stellar atmospheres (e.g. Collet, Asplund \& Th\'evenin 2005; Asplund 2005).

We refer the reader to Asplund (2005) for a review of non-LTE effects in
stellar atmospheres, as well as computational aspects, caveats and
limitations due to the paucity of known collisional rates.

In order to evaluate the differential non-LTE abundance corrections, and
thus compare the abundances here with predicted nucleosynthetic yields,
we have searched the literature for non-LTE abundance corrections appropriate
for the COS\,171 and UMi~28104 atmosphere parameters; this includes the lines
used for the abundances of each element.  Unfortunately, not all lines of all 
elements have been investigated, and we could not find non-LTE calculations
for vanadium.

Incomplete coverage of stellar atmosphere parameter space is a problem which
prevents us from estimating the differential non-LTE abundance corrections
between COS\,171 and UMi~28104 for some elements; for example, the focus on
warm dwarfs in Yan et al. (2015, 2016) prevents reliable differential 
non-LTE abundance corrections for copper.

  In Table~1 we list non-LTE corrections for a variety 
  of species in COS\,171 and UMi~28104; when possible, we have preferred
  results from Bergemann\footnote{http://nlte.mpia.de/} and collaborators, 
  because that group's results
  include more species than other studies.  We make no evaluation of which
  source of non-LTE corrections is more reliable.

  Fortunately, the non-LTE correction differences in Table~1, 
  for most elements are less than 0.10 dex, with the exception of those
  derived from lines of Ti~I, Mn~I, and Co~I.  Since the yield of Mn is
  sensitive to the neutron excess during explosive nucleosynthesis, 
  it is a useful element for diagnostic purposes; yet, unfortunately, 
  Mn abundances suffer from very large non-LTE corrections.

  As Table~1 shows, the Fe~I non-LTE abundance corrections for
  COS\,171 are smaller
  than for UMi~28104, in the same sense as a number of other species; this
  mitigates, to a small extent, the deviation of the non-LTE corrected
  [X/Fe~I] ratios from the LTE values.

\begin{deluxetable}{rlclc}
\label{tab-abunds-final}
\tabletypesize{\scriptsize}
\tablecaption{Adopted LTE Abundances and Non-LTE Corrections}
\tablehead{\\
\colhead{Species} &
\colhead{\rm $\varepsilon$(X)$_{UMi28104}$} &
\colhead{\rm $\Delta\varepsilon_{\rm nlte}$} &
\colhead{\rm $\varepsilon$(X)$_{Cos171}$} &
\colhead{\rm $\Delta\varepsilon_{\rm nlte}$} \\
}
\startdata
C     &    5.97 &      ...  &    6.18 &     ...      \\
O~I   &    7.07 &      0.0  &    6.98 &     0.0      \\
Na~I  &    3.84 &  $-$0.02  &    3.70 &  $+$0.00     \\
Mg~I  &    5.87 &  $+$0.01  &    5.85 &  $-$0.06     \\
Si~I  &    5.68 &  $-$0.02  &    6.01 &  $-$0.02     \\
K~I   &    3.29 &  $-$0.37  &    3.65 &  $-$0.30     \\
Ca~I  &    4.30 &  $+$0.15  &    4.66 &  $+$0.10     \\
Sc~II &    0.98 &      ...  &    1.01 &     ...      \\
Ti~I  &    2.83 &  $+$0.39  &    3.06 &  $+$0.24     \\
Ti~II &    3.08 &      ...  &    3.24 &     ...      \\
V~I   &    1.71 &      ...  &    1.80 &     ...      \\
Cr~I  &    3.22 &  $+$0.04  &    3.98 &  $+$0.04     \\
Mn~I  &    2.83 &  $+$0.42  &    3.21 &  $+$0.15     \\
Fe~I  &    5.37 &  $+$0.11  &    6.10 &  $+$0.05     \\
Co~I  &    2.94 &  $+$0.42  &    3.21 &  $+$0.19     \\
Ni~I  &    4.15 &  $+$ 0.0  &    4.29 &  $+$0.0      \\
Cu~I  &    1.32 &  $+$0.2   &    1.31 &  $+$0.2      \\
Zn~I  &    2.22 &  $-$0.10  &    2.16 &  $-$0.13     \\
Sr~II &    0.57 &      ...  &    0.93 &     ...      \\
Y~II  & \llap{$-$}0.75 &      ...  & \llap{$-$}0.41 &     ...      \\
Ba~II & \llap{$-$}0.99 &      ...  &    0.03 &     ...      \\
La~II & \llap{$-$}1.97 &      ...  & \llap{$-$}0.88 &     ...      \\
Eu~II & \llap{$-$}2.31 &      ...  & \llap{$-$}1.02 &     ...      \\
\enddata
\end{deluxetable}

In Table~2 we present our final adopted LTE abundances
and non-LTE corrections for COS~171 and UMi~28104.  These constitute
the data for comparison with a variety of explosive stellar nucleosynthesis
scenarios.

%
%

\section{Comparison with Nucleosynthesis Predictions}

\begin{deluxetable*}{lccccccc}
\label{tab-SNyields}
\tabletypesize{\scriptsize}
\tablecaption{Diagnostic Element Ratios from Supernovae}
\tablewidth{4.0in}
\tablehead{\\
\colhead{      }  &
\colhead{[C/Fe]}  &
\colhead{[Si/Fe]} &
\colhead{[Ca/Fe]} &
\colhead{[V/Fe]}  &
\colhead{[Cr/Fe]} &
\colhead{[Mn/Fe]} &
\colhead{[Ni/Fe]}   \\
}
\startdata
COS 171 (LTE)   &      $-$0.90 &   $-$0.15   &  $-$0.28 & $-$0.81 & $-$0.31 & $-$0.92   & $-$0.56 \\[14pt]
COS 171 (NLTE)  &      $-$0.95 &   $-$0.22   &  $-$0.23 &   ...   & $-$0.32 & $-$0.82   & $-$0.56 \\[14pt]
\underline{SNIa (DDTa)z=0.01}  &   $-$3.30 &   $-$0.65   &  $-$0.27 & $-$0.76 & $-$0.11 & $-$0.19   & $-$0.07 \\[14pt]
\underline{SNII (WW95)z=0.01Z$_\odot$} & \\
12M$_{\odot}$   &      $-$0.51 &   $-$0.46   &  $-$0.46 & $-$1.11 & $-$0.46 & $-$0.80  &  $+$0.01    \\
15M$_{\odot}$   &      $-$0.34 &   $-$0.14   &  $-$0.13 & $-$0.69 & $-$0.17 & $-$0.60  &  $-$0.06    \\
20M$_{\odot}$   &      $+$0.18 &   $+$0.75   &  $+$0.69 & $+$0.03 & $+$0.49 & $-$0.19  &  $-$1.07    \\[14pt]
\underline{SNII (K06)z=0}    & \\
13M$_{\odot}$   &      $-$0.16 &   $+$0.33   &  $+$0.11 & $-$0.40 & $+$0.08 & $-$0.54  &  $-$0.30   \\
20M$_{\odot}$   &      $-$0.07 &   $+$0.48   &  $+$0.40 & $-$0.42 & $+$0.35 & $-$0.31  &  $-$0.79   \\[14pt]
\underline{SNIa (Sub-Ch)}   & \\
0.88M$_{\odot}$\\
z=0.00025       &      $-$2.16 &   $-$0.19   &  $+$0.21 & $-$1.47 & $+$0.30 & \color{red}$-$1.56  &  \color{red}$-$2.47   \\
z=0.0025        &      $-$2.16 &   $-$0.17   &  $+$0.19 & $-$0.92 & $+$0.30 & \color{red}$-$0.39  &  \color{red}$-$1.76   \\[14pt]
0.97M$_{\odot}$\\
z=0.00025       &      $-$2.61 &   $-$0.45   &  $-$0.05 & $-$1.57 & $+$0.01 & \color{red}$-$1.77  &  \color{red}$-$0.63   \\
z=0.0025        &      $-$2.61 &   $-$0.44   &  $-$0.08 & $-$1.17 & $+$0.01 & \color{red}$-$0.69  &  \color{red}$-$0.63   \\[14pt]
1.06M$_{\odot}$\\
z=0.00025       &      $-$3.00 &   $-$0.71   &  $-$0.24 & $-$1.67 & $-$0.20 & \color{red}$-$1.87  & \color{red} $-$0.60   \\
z=0.0025        &      $-$3.00 &   $-$0.69   &  $-$0.29 & $-$1.37 & $-$0.20 & \color{red}$-$0.90  &  \color{red}$-$0.60   \\[14pt]
\underline{PISN (HW02)z=0}   & \\
242M$_{\odot}$  &      $-$1.14 &   $+$0.25   &  $+$0.13 & $-$1.16 & $+$0.04 & $-$0.74  &  $-$0.29   \\
260M$_{\odot}$  &      $-$1.38 &   $+$0.02   &  $-$0.08 & $-$1.38 & $-$0.15 & $-$0.95  &  $-$0.09   \\[14pt]
\underline{PISN (KYL14)z=0.001}  & \\
250M$_{\odot}$  &      $-$1.60 &   $+$0.31   &  $+$0.46 & $-$0.96 & $+$0.14 & $-$0.56  &  $-$0.39   \\[14pt]
\enddata
\end{deluxetable*}

In this section we compare the chemical composition of COS\,171 with
predicted nucleosynthesis yields from a variety supernova scenarios.
In a first step we simply make a table comparison of the raw LTE and
non-LTE corrected COS\,171 abundances for a few diagnostic elements.
Then for the most promising scenarios, we compare the detailed COS\,171
composition to theoretical yields added to the composition of our standard 
star, UMi~28104, which we have assumed to represent the background composition 
enriched by the COS\,171 progenitor.  This has the advantage
that zero-point measurement errors common to COS\,171 and UMi~28104
cancel-out, giving the smallest measurement uncertainty; however,
differential metallicity effects may still be present.
Finally, we compare the theoretical yields for sub-Chandrasekhar mass SNIa
and Chandrasekhar-mass SNIa deflagration to detonation (DDT) models to the
composition of COS\,171 with 
the UMi~28104 background composition subtracted-out.  While this gives a direct
comparison with predicted yields the uncertainties in the measured abundance
differences can be large, especially for elements with small or zero
enhancement over the background composition.  

All three comparisons
favor a sub-Chandrasekhar mass SNIa as the progenitor to the COS\,171 composition.


\subsection{Comparison of Supernova Yields with the Measured COS\,171 Composition}

In Table~3 we compare the raw LTE and non-LTE corrected [X/Fe]
ratios of a handful of diagnostic elements in COS\,171 with predicted
element yields for various explosive events.  

For core-collapse,
SNII, yields we compare with the predictions of Woosley \& Weaver (1995;
henceforth WW95) and Kobayashi et al. (2006; henceforth K06); for pair 
instability supernovae (PISN) 
predictions, at z=0, we compare with the results of Heger \& Woosley (2002; 
henceforth HW02), while for z=0.001 PISN we use the result of
Kozyreva, Yoon \& Langer (2014).  For predicted SNIa yields of various
masses and metallicities we consider
the Chandrasekhar-mass DDT models of
Badenes et al. (2003, 2008b) and Yamaguchi et al. (2015), and
the sub-Chandrasekhar mass models (henceforth Bravo models)
introduced in Yamaguchi et al. (2015), calculated with a version of
the code described in
Bravo \& Mart\'inez-Pinedo (2012)\footnote{The Bravo sub-Chandrasekhar mass SNIa
yields are tabulated in Tables~\ref{tab-a1}--\ref{tab-a4} in Appendix A of this paper.}.
In particular, we seek predicted yields that could potentially reproduce
the unusually low [Mn/Fe], [Ni/Fe] and [$\alpha$/Fe] ratios seen in
COS\,171, as displayed in Figures~\ref{fig-ch10-alphafe} and
\ref{fig-ch10-iron-peak}.

From Table~3 one can immediately see that massive 
(e.g., $\geq$20 M$_{\odot}$), core-collapse,
SNII fail to reproduce the COS\,171 composition, because of their expected
large $\alpha$-element yields (e.g., Si, Ca); the situation would  be worse
for more massive SNII, especially for [Mg/Fe] which is over-produced in
very massive SNII (e.g. WW95, K06).  Table~3
shows that massive SNII also
over-produce C, V, Cr, and Mn relative to COS\,171.  

Low-mass SNII, such as the 
12M$_{\odot}$ and 13M$_{\odot}$ models of WW95 and K06, respectively,
produce low enough
$\alpha$-elements to be consistent with COS\,171, but they still
significantly over-produce carbon, and nickel relative to COS\,171.  
We note that the yields
for z=0, 10M$_{\odot}$, SNII predicted by Heger \& Woosley (2010) 
indicate [Ni/Fe] ratios close to the solar value, well above the
measured COS\,171 value at $-$0.56 dex.

For PISN the z=0 HW02 260M$_{\odot}$ [X/Fe] yields for C, Si, Ca, V, Cr and
Mn are fairly close to the low values seen in COS\,171, but the predicted
[Ni/Fe] ratio is too high by more than $\sim$0.5 dex.  The
250M$_{\odot}$ z=0.001 PISN model of Kozyreva et al. (2014) gives a very 
poor match to COS\,171, and in particular cannot explain the low X/Fe 
values for Si, Ca, Cr, Mn, and possibly Ni.


For the DDTa Chandrasekhar-mass SNIa model 
Table~3 shows that C/Fe, Si/Fe and Ca/Fe
ratios are low enough to fit COS\,171, but Cr/Fe, Mn/Fe and Ni/Fe
fail significantly.  

The Sub-Chandrasekhar models indicate Mn and Ni abundances
low enough to match the observed COS\,171 Mn/Fe and Ni/Fe ratios
for almost all models presented.  However, the 1.06M$_{\odot}$
sub-Chandrasekhar mass models give [X/Fe] suitably low for all diagnostic
species listed in Table~3.

%
%
%



\subsection{Comparison with Chandrasekhar Mass SNIa}

\begin{figure}[h]
\centering
\vbox{
\includegraphics[width=8.7cm]{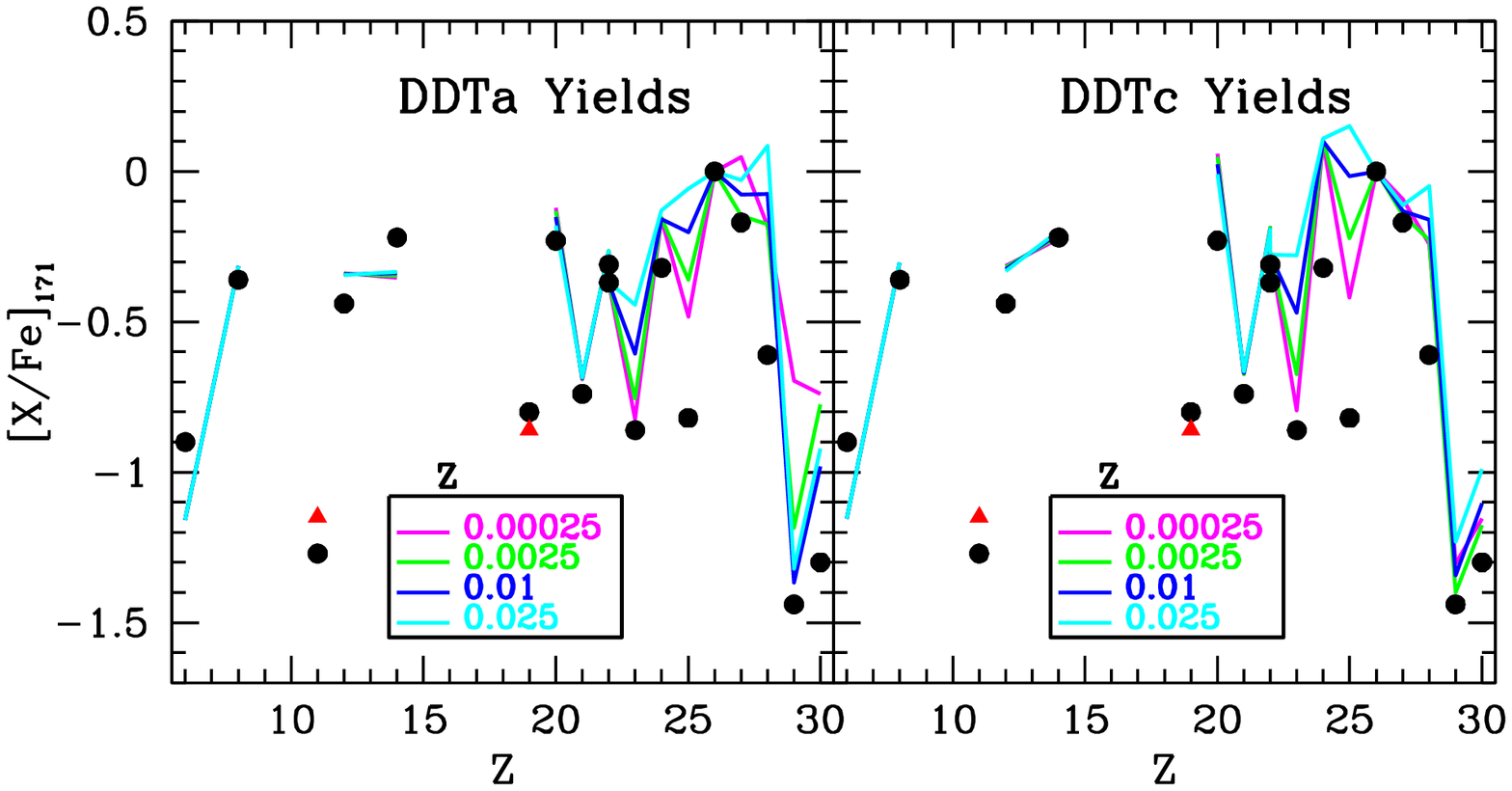}
}
\caption{A comparison of the COS\,171 non-LTE corrected composition
(filled black circles), from C to Zn, with
Bravo (unpublished) Chandrasekhar-mass nucleosynthesis yields, for
various metallicities and two model deflagration to detonation
transition densities (DDTa and DDTc) added to 
the background composition of UMi~28104 (solid lines).  Red lower limit
triangles are shown for Na and K, which were not reported by Bravo, 
indicating the [Na/Fe] and [K/Fe] ratios for zero production of these
elements.
The predicted [Mn/Fe] yields fail to reproduce the low values
seen in COS\,171 and [Ni/Fe] ratios are over-produced in the
models, both by $\sim$0.5 dex, compared to the observations; these
differences are well in excess of the $\sim$0.1 dex measurement
uncertainties.
The [Si/Fe], [Ca/Fe] and [Cr/Fe] ratios are better matched by the DDTa
model.  }
\label{fig-ChSNIa}
\end{figure}

Predicted yields for DDT models added to the UMi~28104 background
composition, at two deflagration to detonation transition densities 
(DDTa and DDTc; e.g., Badenes et al. 2003, 2008b),
and a range of metallicities
are shown in Figure~\ref{fig-ChSNIa}; these give fair agreement with the
measured non-LTE corrected COS\,171 composition.  However, the important 
neutron-rich elements
Mn and Ni are significantly over-produced in these models, well beyond the
measurement uncertainties; Cr/Fe is also poorly fit.  This disagreement could
reasonably be due to the increase in neutron excess during the carbon-burning 
simmering phase (e.g., Piro \& Bildsten 2008).  
Badenes et al. (2003, 2008b)
did not compute Na or K yields, so Figure~\ref{fig-ChSNIa} shows, with red 
lower limits, the [Na/Fe] and [K/Fe] ratios expected for no production in the
COS\,171 progenitor; here the final [Na/Fe] and [K/Fe] ratios are due to the 
UMi~28104 background composition diluted with 0.7 dex of extra Fe.

A range of Chandrasekhar-mass, single degenerate SNIa models, investigated
by Dave et al. (2017), permits some comparison with the composition of COS\,171.  
Although Dave et al. (2017) did not publish their yields, their Figures~4 and 7 
contains yield information for a few elements.  In particular, their 
Figure~4 allows a comparison of the predicted X/Fe mass fractions for Ni, Mn,
and Cr as a function of WD metallicity for all classes of models considered; 
also, their Figure~7 allows an approximate comparison for their standard
deflagration model and their high-density, low C/O ratio, deflagration to
detonation model.

Surprisingly, we found that the Dave et al. (2017) Figure~4 Mn/Fe and
Cr/Fe mass ratios for the zero-metallicity, high-density, gravitationally
confined detonation (z=0, GCD-HIGHDEN) model provides a reasonable match
to the measured Mn/Fe and Cr/Fe mass ratios in COS\,171.  However, the
predicted Ni/Fe mass fractions for all models were factors of 12 
to 60 greater than measured for COS\,171, with the closest match given by 
the GCD-HIGHDEN model.  The STD-DEF and DDT-HIGHDEN-LOWC/CENTRAL models
in Dave et al. (2017) compared badly with COS\,171, with Mn/Fe and
Ni/Fe much larger than observed, by approximately a factor of 10; however,
the Cr/Fe was reasonably well matched by these models.

We conclude that none of the single degenerate models match the
observed composition of COS\,171; in particular, the predicted Mn/Fe and
Ni/Fe ratios appear to rule-out this scenario.

\subsection{Comparison with Pair Instability Supernovae Yields}

\begin{figure}[h]
\centering
\includegraphics[width=8.0cm]{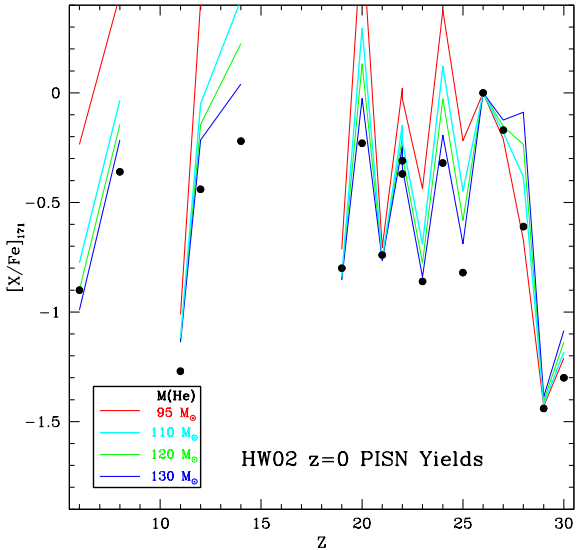}
\caption{A comparison of the HW02 pair instability supernova element yields,
added to the UMi~28104 background composition, with the COS\,171 abundance
distribution (including differential non-LTE corrections).  The closest match 
is obtained for a helium core mass of 130M$_{\odot}$, corresponding
to a total mass of 260M$_{\odot}$; however, the [Ni/Fe] ratio for
that model exceeds the measured value by $\sim$0.5 dex.
}
\label{fig-pisnhw02}
\end{figure}

Interestingly, the predicted PISN element yields, added to the UMi~28104
background composition, for
the 260M$_{\odot}$ z=0 star of HW02 come fairly close to fitting the
measured non-LTE corrected COS\,171 X/Fe ratios for most elements, except
for Ni/Fe, which is too high by more than 0.5 dex.  Given the comparison 
in Figure~\ref{fig-pisnhw02},
an extrapolation of the PISN yields to slightly higher than their maximum
mass might seem to do even better for all elements except Ni.  However,
HW02 stress that the 260 M$_{\odot}$ is strictly the highest possible
z=0 PISN mass; above this limit pair production is unable to provide sufficient 
pressure and the object collapses to a black hole.  

Furthermore, an improved fit with PISN yields cannot be obtained with 
non-zero metallicity PISN events, as indicated by the z=0.001 250M$_{\odot}$
model by Kozyreva et al. (2014), which fails to match the observed 
Si/Fe, Ca/Fe, Cr/Fe, Mn/Fe and Ni/Fe in COS\,171.  Thus, currently predicted
element yields from PISN events do not fit the composition of COS\,171
and we reject this scenario.

\subsection{Comparison with Sub-Chandrasekhar Mass SNIa Models}

\begin{figure}[h!]
\centering
\includegraphics[width=8.7cm]{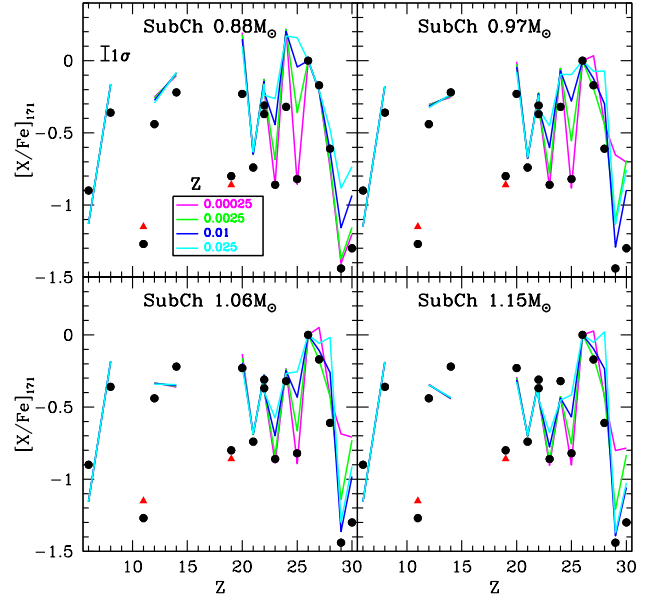}
\caption{A comparison of the COS\,171 non-LTE corrected composition
(filled black circles), from C to Zn, with
Bravo (unpublished) sub-Chandrasekhar mass nucleosynthesis yields, for
various WD masses and primordial metallicities, z, added to the
background composition of UMi~28104 (solid lines).  The predictions 
show that Mn/Fe, V/Fe and Ni/Fe are sensitive to initial metallicity,
while Si/Fe, Ca/Fe and Cr/Fe depend on WD mass.
}
\label{fig-subch-nlte}
\end{figure}

Figure~\ref{fig-subch-nlte} shows excellent overlap between the predicted
sub-Chandrasekhar mass SNIa yields added to the UMi~28104 background
composition and the observed abundances in COS\,171 (corrected for
differential non-LTE effects).  Critically, there is good overlap 
for the important [Mn/Fe] and [Ni/Fe] ratios, but also
for Sc, Ti, V and Co.  In particular, the low [Mn/Fe] and [V/Fe]
ratios seen in COS\,171 are only reproduced in the lowest metallicity
models.

Figure~\ref{fig-subch-nlte} reveals that the Bravo 
sub-Chandrasekhar mass predictions give decreasing [X/Fe] yields of
Si, Ca and Cr with increasing WD mass; conversely, the [Ni/Fe] 
yield increases with increasing WD mass.  These trends suggest that
both the WD mass and metallicity may be constrained using the full array
of element ratios relative to iron.

The low [Ni/Fe], [Cu/Fe] and [Zn/Fe] ratios in COS\,171, while reproduced 
in the the 0.88 M$_{\odot}$, z=0.00025, Bravo model, are not well matched 
with higher WD masses.  However, Si is best reproduced in the 
0.97 M$_{\odot}$ models and Ca, Ti, and Cr abundances in COS\,171 are best 
matched by the 1.06 M$_{\odot}$, z=0.00025, models.  Thus, there is some 
disagreement between the best matching sub-Chandrasekhar WD mass, 
depending on which elements are considered.  While errors in the adopted
differential non-LTE corrections to Ca, Ti and Cr LTE abundances might 
resolve the mass discrepancy, the changes required are at least 0.2 dex
and need to work in the same direction, which seems unlikely.  Alternatively,
an increase in the Fe~I non-LTE correction for UMi~28104 by 0.2 dex would
reduce the progenitor mass indicated by Ca, Ti and Cr, but this would shift 
all the solid curves in Figure~\ref{fig-subch-nlte} down, and the implied
progenitor mass from Si and Ni abundances would move to even lower
values, still out of agreement with Ca, Ti and Cr.

In the Bravo models, Cu and Zn are made in two regions: by alpha-rich
freeze-out in the core and also following carbon burning close to the surface, 
where protons and neutrons are released that subsequently build-up Cu and Zn 
in a series of (n,$\gamma$), (p,$\gamma$) and (n,p) reactions.  Importantly,
the alpha-rich freeze-out cannot occur in the 0.88M$_{\odot}$ model because 
its maximum temperature is insufficient for complete Si-burning, unlike the 
more massive WD models (where maximum temperatures are near 
T$\sim$6~GK).
Furthermore, for the (n,$\gamma$) and (n,p) source of Cu and Zn near the
surface, the controlling neutron-excess is determined by the original stellar 
metallicity in the sub-Chandrasekhar mass models.  Therefore, the observed 
low abundances of Cu and Zn in COS\,171 suggests both low-metallicity 
(z=0.00025) and low WD mass (0.88M$_{\odot}$).

\begin{figure}[h]
\centering
\includegraphics[width=8.0cm]{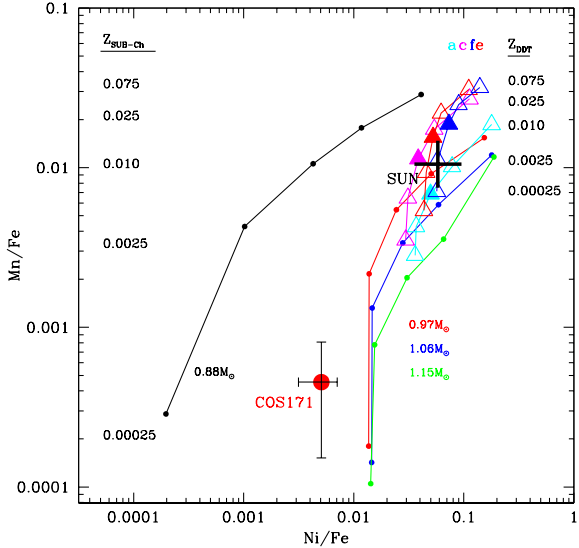}
\caption{Estimated Ni/Fe and Mn/Fe mass ratios for the
COS\,171 progenitor compared to DDT and sub-Chandrasekhar mass models.
Triangles: DDT models with
different deflagration to detonation transition densities (a, c, e, f)
and metallicities; filled triangles indicate Z$_{DDT}$=0.010, 
slightly above the effective value arising from simmering.
Small filled circles connected by lines: various sub-Chandrasekhar 
mass models, for different metallicities.
Large filled red circle: COS\,171 with background composition of UMi~28104
subtracted.  Large black cross:  solar value.
 Note that COS\,171 lies in a low-metallicity
Mn/Fe region, much lower than possible by DDT models (due to 
simmering increase of $\eta$); COS\,171 also has low Ni/Fe, in
a region sensitive to WD mass, due to the reduced role of 
alpha-rich freeze-out nucleosynthesis.  At higher metallicity, the
Ni/Fe ratio and Mn/Fe ratios increase due to neutron-excess
dependent nucleosynthesis.
}
\label{fig-subch_nifemnfe}
\end{figure}

\begin{figure}[h]
\centering
\includegraphics[width=8.0cm]{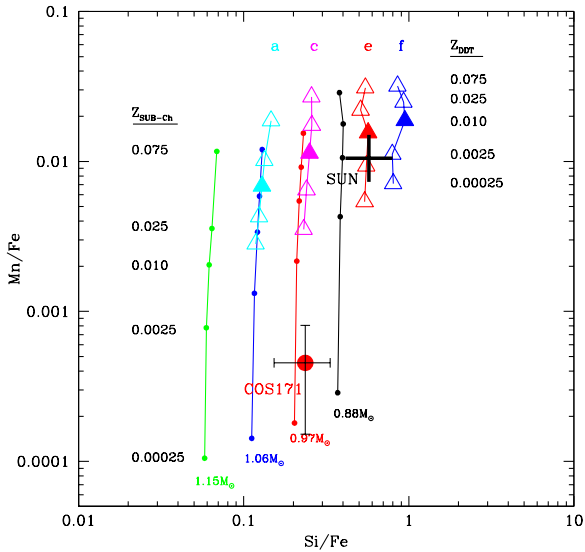}
\caption{Estimated Si/Fe and Mn/Fe mass ratios for the
COS\,171 progenitor compared to DDT and sub-Chandrasekhar mass models.
Symbols are the same as Figure~\ref{fig-subch_nifemnfe}.
The Si/Fe ratio increases with decreasing WD mass, likely to
greater fraction of the core that experiences incomplete Si burning.
The Si/Fe ratio indicates a WD mass near 0.94 M$_{\odot}$, consistent
with the value from Ni/Fe in Figure~\ref{fig-subch_nifemnfe}.
Note the clean separation between WD mass and metallicity indicated
by Si/Fe and Mn/Fe in this plot.
}
\label{fig-subch_sifemnfe}
\end{figure}

Direct comparison of the background-subtracted COS\,171 element mass ratios
with predicted sub-Chandrasekhar mass yields are presented in
Figures~\ref{fig-subch_nifemnfe} and \ref{fig-subch_sifemnfe} for Mn/Fe
as a function of Ni/Fe and Si/Fe, respectively.
These two figures employ abundance differences, by number, of
COS\,171 minus UMi~28104, compared to the predictions of Bravo (unpublished).  
Because these mass ratios are based on true abundance differences, the
measurement uncertainties lead to large uncertainties 
for elements with similar abundances in the two stars; this is
despite the small abundance measurement uncertainties, near 0.1 dex, seen in
Figure~\ref{fig-subch-nlte}.  

We estimated the relative abundance difference uncertainties in 
Figure~\ref{fig-subch_nifemnfe} and \ref{fig-subch_sifemnfe} arising
from scatter in CH10 measurements, including
covariances and random excitation temperature errors, and assuming a 0.03 dex
loggf scatter.  Accordingly, we found 1$\sigma$ uncertainties 
of 0.070 dex, 0.047 dex, and 0.089 dex for for Mn/Fe, Ni/Fe, and Si/Fe,
respectively. Thus, the small total relative abundance uncertainties
translate into the substantial mass ratio uncertainties shown in
Figures~\ref{fig-subch_nifemnfe} and \ref{fig-subch_sifemnfe}.

In Figure~\ref{fig-subch_nifemnfe} the low Mn/Fe mass ratio indicates a low
metallicity for the COS\,171 progenitor, within 1$\sigma$ of the lowest 
metallicity in the Bravo models, at z=0.00025, corresponding to
[Fe/H]=$-$1.7 dex.  Thus, the COS\,171 Mn/Fe mass ratio is consistent with
a low metallicity sub-Chandrasekhar mass SNIa, and lies well below any of 
the DDT model predictions.  

The COS\,171 Ni/Fe mass ratio in Figure~\ref{fig-subch_nifemnfe} also lies
well below the DDT models, and below most of the sub-Chandrasekhar mass 
model predictions, and indicates a low WD mass between 0.97 and 
0.88 M$_{\odot}$.
We note that for lowest WD masses considered by Bravo, the maximum core
temperatures are insufficient to drive the alpha-rich freeze-out, which 
is an important source of Ni, as well as for Cu and Zn.  In the Bravo 
models, the reduced Ni production from the alpha-rich freeze-out is likely 
the reason why the predicted Ni/Fe
ratio declines steeply below 0.97 M$_{\odot}$ and at low metallicity.  
However, at high metallicity, there is
an increase in Ni/Fe, even for the 0.88 M$_{\odot}$ models; this is
likely due to increased neutron excess for higher metallicity 
sub-Chandrasekhar mass SNIa.

The importance of the alpha-rich freeze-out in the production of Ni,
Cu and Zn leads to a sensitivity of these elements to the WD progenitor
mass as well as metallicity; indeed, the measured low Ni/Fe, Cu/Fe and Zn/Fe 
ratios for COS\,171 suggest that the progenitor was likely low-mass,
(a linear interpolation suggests 0.95 M$_{\odot}$) with a reduced or
no alpha-rich freezout and very low metallicity.  The low 
metallicity is supported by the measured low Mn/Fe and V/Fe ratios.

Figure~\ref{fig-subch_sifemnfe} also shows that the COS\,171 composition lies
outside the range of Chandrasekhar-mass SNIa in Table~3.
Linear interpolation over Si/Fe for the sub-Chandrasekhar mass models in
Figure~\ref{fig-subch_sifemnfe} suggests
a mass of 0.94 M$_{\odot}$, consistent with that indicated by the Ni/Fe ratios.
Contrary to the case for Ni/Fe, however, the Si/Fe ratio increases with
decreasing WD mass; this is likely due to the relatively low temperatures of 
the low mass models, resulting in incomplete Si burning and larger amounts of 
unburnt Si remaining.

We may compare the subtracted COS\,171 minus UMi~28104 element mass
ratios with the sub-Chandrasekhar mass SNIa mass ratio predictions
of Shen et al. (2017), appearing in their figures 8--11. Our
subtracted Mn/Fe and Ni/Fe mass ratios for COS\,171 lie close to
the 0.9 M$_{\odot}$, z=0, result of Shen et al. (2017) in their
Figure~8, but suggest a mass near 0.97 M$_{\odot}$ in their Figure~9,
(both adopt a C/O mass ratio of 50/50).  On the other hand, the
Shen et al. (2017) predictions for Cr/Fe mass ratios, appearing in their
Figure~11, barely overlaps with the measured value for COS\,171, but is
best matched in their models with a WD mass of 1.1 M$_{\odot}$.
As mentioned previously, Cr/Fe mass ratios predicted by Bravo suggest
a match to COS\,171 near 1.06 M$_{\odot}$.  Thus, both sets of
sub-Chandrasekhar mass models suffer the same differences with our
best estimate for the element ratios in COS\,171.

We are encouraged by the agreement between Shen et al. (2017) and 
Bravo predictions for sub-Chandrasekhar mass SNIa element yields of
Si, Mn, Fe, and Ni; of particular significance is the implied low
sub-Chandrasekhar mass SNIa, which is also supported by the large
deficiencies of alpha-rich freeze-out elements Ni, Cu, and Zn.

However, the higher mass WD progenitor indicated from the Ca/Fe, Ti/Fe 
and Cr/Fe ratios suggest that details of the actual SNIa event differed
slightly from the models.

\section{Chemical Evolution of UMi and COS\,171}

Here we discuss possible chemical evolution scenarios to explain the
unusual chemical composition of COS\,171 in UMi.  As mentioned previously, 
it is clear that the composition of the slightly more metal-poor stars in 
UMi were not on the chemical path to COS\,171.  However, it appears that 
COS\,171 resulted from the addition of iron-peak and r-process material
to the composition of UMi stars near [Fe/H]=$-$2 dex, such as UMi~28104.

\subsection{Contamination of an existing star in the proximity of a single SNIa}

It is possible that the unusual composition of COS\,171, dominated by
a sub-Chandrasekhar mass SNIa, may have been due to direct accretion 
of ejecta from a nearby supernova.  In this scenario, COS\,171 would have
been the outer star in a triple system orbiting a pair of merging low-mass WDs.  
Assuming an envelope mass for COS\,171 of 0.4 M$_{\odot}$, then at
[Fe/H]=$-$1.35 dex, it contains 2.8$\times$10$^{-5}$M$_{\odot}$ of iron, of
which 6.3$\times$10$^{-6}$M$_{\odot}$ was due to the pre-existing,
[Fe/H]$\sim$$-$2 dex, material.  If 0.5 M$_{\odot}$ of Fe was produced by
the SNIa event, then a fraction of only 4.3$\times$10$^{-5}$ of the
SNIa iron was captured by COS\,171.  If COS\,171 was a main sequence star at the
time of the accretion, with roughly solar radius, and assuming that the accretion
radius was equal to the physical radius, then COS\,171 must have been separated 
from the SNIa event by $\sim$76 R$_{\odot}$, roughly 0.4 AU, in order to reach
[Fe/H]=$-$1.35 dex.  This distance is smaller than the radius of the RGB star
phase of the WD progenitors, so seems unlikely.

On the other hand, if COS\,171 was an RGB star, of 20R$_{\odot}$
radius, at the time of accretion then the separation from the SNIa event
would have been approximately 7 AU, which is larger than the maximum radius
of the AGB phase (roughly 1.5 AU) of the progenitor WDs.
These separations seem small considering that merger of two WDs would be
required for the SNIa event in an even smaller region, and must also have
accommodated the previous RGB and AGB phases for both WD progenitors.  Thus, 
if this mechanism did occur, there may well have been a reduction in the size
of the COS\,171 orbit prior to the accretion and supernova event.  Furthermore,
in the RGB accretion scenario the SNIa event must have occurred relatively
recently.  Given these constraints, we do not favor this scenario.

\subsection{Contamination of a typical molecular cloud by many SNIa}

In order to reach the chemical composition of COS\,171 by pollution of a 
giant molecular cloud, near 10$^6$ M$_{\odot}$, ejecta from approximately
100 SNIa are required.  The difficulty with such a scenario is that with many 
SNe one might expect to see an averaging of element yields from other nucleosynthesis
events, in particular SNII and Chandrasekhar-mass SNIa, in which case the unusual
abundance ratios seen in COS\,171 might not be expected.  For such a scenario a
systematic modulation of the polluting supernova ejecta appears to be necessary in 
order to obtain the COS\,171 composition, dominated by sub-Chandrasekhar mass SNIa 
ejecta.

To circumvent this problem, one could propose that only sub-Chandrasekhar mass
SNIa occurred, during a late phase of chemical enrichment in UMi, perhaps from
the last epoch of iron-peak enrichment by the longest-lived progenitors, which 
led to the SNIa in UMi dominated by the merger of sub-Chandrasekhar mass WD stars.

Evidence for large-scale enrichment by low-metallicity sub-Chandrasekhar mass SNIa
in other dwarf galaxies could be construed, qualitatively, from the deficiencies
of V, Mn, Co, Ni, Cu and Zn, relative to Fe, in the Sagittarius dwarf galaxy
by Hasselquist et al. (2017), Sbordone et al. (2007), and  McWilliam et al. (2003),
Ni deficiencies in LMC stars found by Van der Swaelmen et al. (2013),
and Ni deficiencies in the Fornax dwarf galaxy by Letarte et al. (2010).
However, the element deficiencies in these other dwarf galaxies are much less
extreme than those seen in COS\,171, as if mixing with a range of
supernova yields occurred in these systems, unlike COS\,171.

Detailed investigation into the less extreme deficiencies, seen elsewhere, is
required before they are taken as solid evidence of sub-Chandrasekhar mass SNIa
nucleosynthesis; and, they likely require dilution with element yields from other
supernova types.  Thus, it is possible that sub-Chandrasekhar mass SNIa
nucleosynthesis is common among dwarf galaxies, and therefore that a
sub-Chandrasekhar mass SNIa dominated phase might have occurred in UMi, but it
is not yet proven.

In this pollution scenario, we would expect many other stars sharing the chemical 
composition of COS\,171 to be present in UMi; such objects may yet be found among
the lower luminosity UMi stars.  
Indeed, the frequency of stars in UMi that share the chemical composition of COS\,171
would provide strong constraints on the three scenarios outlined here.

\subsection{Contamination of interstellar gas by a single SNIa}

One possibility is that the iron-peak composition of COS\,171 resulted from 
pollution of interstellar gas by a single SNIa event.  
Since individual SNIa events typically yield approximately 0.5 M$_{\odot}$ of 
iron, this must be diluted with $\sim$9,000 M$_{\odot}$ of [Fe/H]=$-$2 dex gas,
in order to reach the COS\,171 metallicity, at [Fe/H]=$-$1.35 dex.  This
amount of gas dilution corresponds to a relatively low-mass molecular cloud.

Detailed modelling of the evolution of supernova remnants (henceforth SNR),
by Chevalier (1974), show that the amount of interstellar material (ISM) 
swept-up by a SNR depends on many factors, including the energy of the
explosion, the density of the ISM, the strength of the local magnetic fields, 
and cooling from metals and grains.  

Equations describing SNR growth, in Chevalier (1974), indicate that at an
ISM density of 1.0 cm$^{-3}$ (characteristic of the warm ISM) the typical
SNR shell velocity drops to the dispersion of interstellar clouds at a 
radius near 44pc, indicating a swept-up mass of 10$^4$ M$_{\odot}$.
This is completely consistent with the detailed SNR treatment 
of Cioffi et al. (1988), also including metal-dependent cooling, which gives a
radius of 42pc, again with an implied swept-up mass near 10$^4$ M$_{\odot}$.

A more recent calculation, by Asvarov (2014), is also consistent with these
results, and estimated the largest SNR radius of 45pc for a remnant expanding
into an ISM with uniform density of 1.0 cm$^{-3}$, over 4$\times$10$^5$ yr.  A
radius of 34pc was found when the magnetic field pressure was
increased by a factor of 4.  Notably, Asvarov (2014) employed a relatively small
critical velocity, at Mac 2.  

These predictions are consistent with the distribution of SNR sizes in nearby
galaxies, for example as found by Badenes et al. (2010), who found a sharp
cut-off at a radius near 30pc, and the largest SNR radii near 60pc.

The formation of molecular clouds out of enriched warm interstellar gas
is the subject of on-going research (e.g., see V\'azquez-Semadeni et al. 2007); 
however, it appears that molecular clouds are short-lived objects within an 
on-going, equilibrium, process of rapid formation by gravitational instability
and equally rapid disruption by stellar feedback
(e.g., Mac Low, Burkert, \& Ib\'a\~nez-Mej\'ia 2017). 

%

Given these considerations, it seems reasonable that as UMi ran-out of gas 
and star formation in the galaxy drew to a close, that a single, stochastic,
event might
have enriched 10$^4$ M$_{\odot}$ of warm interstellar gas, all or part of which
subsequently formed into a star-forming molecular cloud, producing the last few
stars in UMi, including COS\,171.  Such an event might be more
likely to result from a long-lived progenitor, delayed from a previous star
formation epoch, such as a sub-Chandrasekhar-mass SNIa.  
Given the requirement for mixing of the SNIa ejecta with $\sim$10$^4$ M$_{\odot}$ 
of hydrogen, in order to produce COS\,171, and that this is the natural outcome for
a SNR mixing into warm interstellar gas, this is our favored scenario for the
origin of COS\,171.

\section{Summary}

We have investigated the highly unusual chemical composition, found by CH10,
of star COS\,171 in the UMi dwarf galaxy.  We confirm the stellar
atmosphere parameters and LTE iron and other element abundances found by
CH10, based on their published EWs.  However, for oxygen abundances we
employ the 6300\AA\ [O~I] line, from the CH10 spectra, corrected for
telluric contamination.  We also revise the potassium abundance in
COS\,171 down by 0.37 dex, based on an EW re-measurement of the 
somewhat saturated K~I line at 7699.0\AA\ in the CH10 spectrum.

The composition of COS\,171 is unlike any MW halo star, with uniquely
low X/Fe ratios for O, Mg, Si, Ca, Ti, Sc, V, Mn, Ni, Cu and Zn.
Abundance ratio plots reveal that this unusual chemical composition 
seems to result from the addition of $\sim$0.7 dex of
iron-peak material to a pre-existing composition, near [Fe/H]=$-$2.05 dex.
Other UMi stars, slightly more metal-poor than COS\,171, do not 
share the same iron-peak chemical locus, although an r-process enrichment 
in those stars is also seen in COS\,171.  The r-process enrichment appears
to be disconnected and separate from the iron-peak peculiarity of COS\,171.

We adopt star UMi~28104, with [Fe/H]=$-$2.08 dex, as a standard for comparison
with COS\,171, since UMi~28104 has a metallicity close to our estimate of the
pre-existing [Fe/H] prior to COS\,171.  The similarity of the atmosphere
parameters of COS\,171 and UMi~28104 also results in a mitigation of systematic
measurement errors, from various effects, in a differential comparison of 
abundance ratios.

Where possible, we have applied non-LTE corrections for the abundance
ratios in COS\,171 relative to UMi~28104, based on a variety of currently
available non-LTE studies.  However, no non-LTE corrections were available
for vanadium.

An abundance ratio plot, of COS\,171 over UMi~28104, shows the 0.7 dex Fe 
enhancement, and clear enhancements of Si, Ca, Cr in COS\,171; mild or
zero enhancements of C, Ti, Mn, Ni and Co are present.  However, O, Na, Mg,
K, Sc, V, Cu and Zn show no evidence of production between UMi~28104 and
COS\,171.

We have compared the composition of COS\,171 with a variety of supernova
nucleosynthesis predictions for a range of metallicities: low and high-mass
core-collapse SNII; Chandrasekhar-mass SNIa; sub-Chandrasekhar mass SNIa; and
pair instability supernovae.  We find, in particular, that the Mn/Fe and
Ni/Fe abundance ratios in COS\,171 can only be reproduced in low-metallicity 
sub-Chandrasekhar mass SNIa nucleosynthesis.  Chandrasekhar-mass SNIa fail
to reproduce the low Mn/Fe ratios, due to pre-explosive simmering that
increases the neutron excess and the yield of neutron-rich species like Mn.
Furthermore, the low Ni/Fe, Cu/Fe and Zn/Fe suggest an absence of alpha-rich
freeze-out nucleosynthesis, which indicates a relatively low-mass 
sub-Chandrasekhar mass SNIa. Our best estimate for the mass of the WD SNIa 
progenitor to COS\,171, based on the predictions
of Bravo (unpublished), is 0.94 M$_{\odot}$.

We conclude that COS\,171 shows direct evidence of sub-Chandrasekhar mass
SNIa nucleosynthesis.

We find that in order to reproduce the COS\,171 metallicity by adding a 
a single SNIa event into the pre-existing UMi composition (near
[Fe/H]=$-$2.05 dex), dilution with approximately 10$^4$ M$_{\odot}$ of
hydrogen is required.  Detailed calculations show that supernova remnants 
expanding into a warm interstellar medium, with a density near 1.0 cm$^{-3}$,
must mix with 10$^4$ M$_{\odot}$ for the expansion velocity to reduce to
the observed velocity dispersion of interstellar clouds, and could naturally
explain the measured [Fe/H] of COS\,171.  

%

In our favored scenario for the chemical evolution of UMi and formation
of COS\,171, as UMi ran-out of gas and star formation in the galaxy drew to 
a close, a single, stochastic, low-metallicity sub-Chandrasekhar mass SNIa,
with WD mass near 0.94 M$_{\odot}$, enriched roughly 10$^4$ M$_{\odot}$ of
warm interstellar gas, all or part of which subsequently formed into a
star-forming molecular cloud, resulting in the last few stars in UMi,
including COS\,171.

%
%
%
%
%
%

\acknowledgements
Andrew McWilliam would like to thank Alex Heger and Boaz Katz for useful
conversations.
Carles Badenes acknowledges support from grant NASA ADAP NNX15AM03G S01.
Eduardo Bravo acknowledges funding from the MINECO-FEDER grant AYA2015-63588-P.

\appendix
\centerline{A. Sub-Chandrasekhar Mass SNIa Nucleosynthesis Yields}
\vskip0.3cm
In Tables~\ref{tab-a1}--\ref{tab-a4} below,
we give the element yields for sub-Chandrasekhar mass SNIa from
the calculations of Bravo (unpublished), introduced in Yamaguchi et al. (2015),
for a range of masses and metallicities.
\vfill\eject

\vskip0.3cm
\begin{table}[h]
\tabletypesize{\scriptsize}
\centering
\caption{0.88~M$_{\odot}$ Sub-Chandrasekhar SNIa Yields (in M$_{\odot}$)}
\tablewidth{4.5in}
\begin{tabular}{lccccc}
\hline\hline
Elem./Z & 0.00025 &    0.0025     &     0.01      &     0.025     &   0.075  \\[4pt]
\hline
C  &  5.164e-03   &   5.167e-03   &   5.170e-03   &   5.148e-03   &   5.028e-03   \\
N  &  6.051e-06   &   5.441e-06   &   4.852e-06   &   4.400e-06   &   4.102e-06   \\
O  &  1.041e-01   &   1.068e-01   &   1.112e-01   &   1.153e-01   &   1.189e-01   \\
Ne &  3.609e-03   &   3.547e-03   &   3.496e-03   &   3.498e-03   &   3.702e-03   \\
Mg &  3.114e-02   &   2.828e-02   &   2.266e-02   &   1.654e-02   &   9.179e-03   \\
Si &  1.531e-01   &   1.572e-01   &   1.609e-01   &   1.626e-01   &   1.597e-01   \\
S  &  9.769e-02   &   9.780e-02   &   9.874e-02   &   9.867e-02   &   8.860e-02   \\
Ar &  2.650e-02   &   2.576e-02   &   2.458e-02   &   2.280e-02   &   1.786e-02   \\
Ca &  3.333e-02   &   3.162e-02   &   2.879e-02   &   2.489e-02   &   1.613e-02   \\
Sc &  8.913e-07   &   9.088e-07   &   9.374e-07   &   1.047e-06   &   2.029e-06   \\
Ti &  4.380e-04   &   4.271e-04   &   3.981e-04   &   3.588e-04   &   2.702e-04   \\
V  &  4.140e-06   &   1.441e-05   &   3.769e-05   &   6.605e-05   &   1.105e-04   \\
Cr &  1.188e-02   &   1.173e-02   &   1.122e-02   &   1.043e-02   &   8.422e-03   \\
Mn &  1.184e-04   &   1.751e-03   &   4.293e-03   &   7.216e-03   &   1.207e-02   \\
Fe &  4.128e-01   &   4.090e-01   &   4.057e-01   &   4.060e-01   &   4.201e-01   \\
Co &  9.950e-07   &   4.802e-06   &   3.575e-06   &   7.092e-06   &   8.623e-05   \\
Ni &  8.124e-05   &   4.168e-04   &   1.735e-03   &   4.773e-03   &   1.720e-02   \\
Cu &  8.800e-07   &   1.859e-06   &   1.197e-05   &   3.472e-05   &   8.641e-05   \\
Zn &  8.800e-07   &   7.718e-06   &   5.012e-05   &   1.120e-04   &   9.344e-05   \\
\hline
\end{tabular}
\label{tab-a1}
\end{table}
\begin{table}[h]
\tabletypesize{\scriptsize}
\centering
\caption{0.97~M$_{\odot}$ Sub-Chandrasekhar SNIa Yields (in M$_{\odot}$)}
\tablewidth{4.5in}
\begin{tabular}{lccccc}
\hline\hline
Elem./Z & 0.00025 &    0.0025     &     0.01      &     0.025     &   0.075  \\[4pt]
\hline
C  &  2.673e-03   &   2.673e-03   &   2.668e-03   &   2.646e-03   &   2.563e-03   \\
N  &  5.214e-06   &   4.515e-06   &   3.761e-06   &   3.054e-06   &   2.355e-06   \\
O  &  7.320e-02   &   7.532e-02   &   7.791e-02   &   8.045e-02   &   8.264e-02   \\
Ne &  1.414e-03   &   1.387e-03   &   1.365e-03   &   1.364e-03   &   1.463e-03   \\
Mg &  2.044e-02   &   1.830e-02   &   1.425e-02   &   1.003e-02   &   5.308e-03   \\
Si &  1.219e-01   &   1.254e-01   &   1.280e-01   &   1.291e-01   &   1.256e-01   \\
S  &  7.818e-02   &   7.822e-02   &   7.912e-02   &   7.907e-02   &   7.123e-02   \\
Ar &  2.118e-02   &   2.055e-02   &   1.980e-02   &   1.852e-02   &   1.490e-02   \\
Ca &  2.635e-02   &   2.484e-02   &   2.293e-02   &   1.999e-02   &   1.331e-02   \\
Sc &  9.746e-07   &   9.997e-07   &   1.007e-06   &   1.061e-06   &   1.495e-06   \\
Ti &  3.605e-04   &   3.592e-04   &   3.375e-04   &   3.037e-04   &   2.307e-04   \\
V  &  4.806e-06   &   1.185e-05   &   3.057e-05   &   5.261e-05   &   8.691e-05   \\
Cr &  8.849e-03   &   8.751e-03   &   8.337e-03   &   7.665e-03   &   6.192e-03   \\
Mn &  1.081e-04   &   1.294e-03   &   3.218e-03   &   5.301e-03   &   8.423e-03   \\
Fe &  5.995e-01   &   5.978e-01   &   5.903e-01   &   5.787e-01   &   5.464e-01   \\
Co &  9.889e-04   &   5.310e-05   &   2.814e-04   &   4.666e-04   &   8.119e-04   \\
Ni &  8.199e-03   &   8.242e-03   &   1.442e-02   &   2.923e-02   &   8.383e-02   \\
Cu &  1.015e-04   &   2.195e-05   &   7.503e-06   &   1.934e-05   &   4.863e-05   \\
Zn &  1.846e-04   &   1.921e-04   &   8.754e-05   &   1.471e-04   &   1.906e-04   \\
\hline
\end{tabular}
\label{tab-a2}
\end{table}
\vbox{
\vskip6.0cm}
\begin{table}[h]
\tabletypesize{\scriptsize}
\centering
\caption{1.06~M$_{\odot}$ Sub-Chandrasekhar SNIa Yields (in M$_{\odot}$)}
\tablewidth{4.5in}
\begin{tabular}{lccccc}
\hline\hline
Elem./Z & 0.00025 &    0.0025     &     0.01      &     0.025     &   0.075  \\[4pt]
\hline
C  & 1.453e-03   &  1.445e-03   &  1.432e-03   &  1.409e-03   &  1.346e-03   \\
N  & 4.559e-06   &  3.716e-06   &  2.895e-06   &  2.281e-06   &  1.678e-06   \\
O  & 4.123e-02   &  4.443e-02   &  4.779e-02   &  4.972e-02   &  5.100e-02   \\
Ne & 6.186e-04   &  6.101e-04   &  6.010e-04   &  6.002e-04   &  6.494e-04   \\
Mg & 1.017e-02   &  9.263e-03   &  7.123e-03   &  4.852e-03   &  2.452e-03   \\
Si & 8.820e-02   &  9.143e-02   &  9.410e-02   &  9.432e-02   &  9.004e-02   \\
S  & 6.102e-02   &  5.997e-02   &  5.939e-02   &  5.866e-02   &  5.256e-02   \\
Ar & 1.732e-02   &  1.623e-02   &  1.503e-02   &  1.409e-02   &  1.187e-02   \\
Ca & 2.231e-02   &  2.000e-02   &  1.732e-02   &  1.510e-02   &  1.038e-02   \\
Sc & 1.104e-06   &  1.091e-06   &  1.110e-06   &  1.137e-06   &  1.317e-06   \\
Ti & 3.003e-04   &  2.994e-04   &  2.827e-04   &  2.570e-04   &  1.961e-04   \\
V  & 4.999e-06   &  9.989e-06   &  2.554e-05   &  4.399e-05   &  7.216e-05   \\
Cr & 7.209e-03   &  7.170e-03   &  6.847e-03   &  6.334e-03   &  5.179e-03   \\
Mn & 1.123e-04   &  1.041e-03   &  2.625e-03   &  4.440e-03   &  8.340e-03   \\
Fe & 7.882e-01   &  7.874e-01   &  7.767e-01   &  7.570e-01   &  6.952e-01   \\
Co & 1.417e-03   &  1.491e-04   &  4.509e-04   &  6.932e-04   &  7.318e-04   \\
Ni & 1.145e-02   &  1.158e-02   &  2.164e-02   &  4.455e-02   &  1.244e-01   \\
Cu & 1.220e-04   &  2.539e-05   &  4.476e-06   &  9.157e-06   &  2.098e-05   \\
Zn & 2.385e-04   &  2.216e-04   &  7.440e-05   &  1.039e-04   &  1.256e-04   \\
\hline
\end{tabular}
\label{tab-a3}
\end{table}
\begin{table}[h]
\tabletypesize{\scriptsize}
\centering
\caption{1.15~M$_{\odot}$ Sub-Chandrasekhar SNIa Yields (in M$_{\odot}$)}
\tablewidth{4.5in}
\begin{tabular}{lccccc}
\hline\hline
Elem./Z & 0.00025 &    0.0025     &     0.01      &     0.025     &   0.075  \\[4pt]
\hline
C  & 7.354e-04   &  7.340e-04   &  7.319e-04   &  7.233e-04   &  6.917e-04   \\
N  & 3.312e-06   &  2.970e-06   &  2.415e-06   &  1.918e-06   &  1.430e-06   \\
O  & 2.028e-02   &  2.089e-02   &  2.180e-02   &  2.248e-02   &  2.301e-02   \\
Ne & 2.353e-04   &  2.320e-04   &  2.311e-04   &  2.340e-04   &  2.651e-04   \\
Mg & 4.516e-03   &  3.949e-03   &  2.890e-03   &  1.875e-03   &  9.078e-04   \\
Si & 5.642e-02   &  5.769e-02   &  5.911e-02   &  5.960e-02   &  5.784e-02   \\
S  & 3.969e-02   &  3.966e-02   &  3.984e-02   &  3.929e-02   &  3.444e-02   \\
Ar & 1.106e-02   &  1.081e-02   &  1.040e-02   &  9.743e-03   &  7.778e-03   \\
Ca & 1.374e-02   &  1.317e-02   &  1.216e-02   &  1.072e-02   &  7.203e-03   \\
Sc & 1.204e-06   &  1.162e-06   &  1.179e-06   &  1.189e-06   &  1.258e-06   \\
Ti & 2.178e-04   &  2.196e-04   &  2.073e-04   &  1.849e-04   &  1.317e-04   \\
V  & 4.644e-06   &  7.757e-06   &  1.984e-05   &  3.393e-05   &  5.454e-05   \\
Cr & 5.163e-03   &  5.171e-03   &  4.959e-03   &  4.579e-03   &  3.930e-03   \\
Mn & 1.021e-04   &  7.557e-04   &  1.957e-03   &  3.313e-03   &  9.813e-03   \\
Fe & 9.726e-01   &  9.718e-01   &  9.570e-01   &  9.280e-01   &  8.413e-01   \\
Co & 1.525e-03   &  2.890e-04   &  6.190e-04   &  8.838e-04   &  6.237e-04   \\
Ni & 1.387e-02   &  1.503e-02   &  2.916e-02   &  6.072e-02   &  1.579e-01   \\
Cu & 1.076e-04   &  2.199e-05   &  2.902e-06   &  4.689e-06   &  8.404e-06   \\
Zn & 2.257e-04   &  1.878e-04   &  5.592e-05   &  6.754e-05   &  7.518e-05   \\
\hline
\end{tabular}
\label{tab-a4}
\end{table}

{}

\end{document}